\documentclass[iop,apj]{emulateapj}
\usepackage{epsfig}
\usepackage{amsmath}
\usepackage{natbib}

\slugcomment{Draft version Jan 6}

\shorttitle{Searching for circumplanetary around LkCa~15}
\shortauthors{Isella A. et al.}

\begin{document}

%% LaTeX will automatically break titles if they run longer than
%% one line. However, you may use \\ to force a line break if
%% you desire.

\title{Searching for circumplanetary disks around LkCa~15} 
 
%% Use \author, \affil, and the \and command to format
%% author and affiliation information.
%% Note that \email has replaced the old \authoremail command
%% from AASTeX v4.0. You can use \email to mark an email address
%% anywhere in the paper, not just in the front matter.
%% As in the title, use \\ to force line breaks.

\author{Andrea Isella}
\affil{Department of Astronomy, California Institute of Technology, MC 249-17, Pasadena, CA 91125, USA}
\email{isella@astro.caltech.edu}

\author{Claire J. Chandler}
\affil{National Radio Astronomy Observatory, PO Box O, Socorro, NM 87801, USA}

\author{John M. Carpenter}
\affil{Department of Astronomy, California Institute of Technology, MC 249-17, Pasadena, CA 91125, USA}

\author{Laura M. P\'erez}
\affil{National Radio Astronomy Observatory, PO Box O, Socorro, NM 87801, USA}

\author{Luca Ricci}
\affil{Department of Astronomy, California Institute of Technology, MC 249-17, Pasadena, CA 91125, USA}

%% Mark off your abstract in the ``abstract'' environment. In the manuscript
%% style, abstract will output a Received/Accepted line after the
%% title and affiliation information. No date will appear since the author
%% does not have this information. The dates will be filled in by the
%% editorial office after submission.

\begin{abstract}
We present Karl G. Jansky Very Large Array (VLA) observations of the 7~mm continuum emission from the 
disk  surrounding the  young star LkCa~15. The observations achieve an angular resolution of 70~mas 
and spatially resolve the circumstellar emission on a spatial scale of 9~AU. The continuum emission traces 
a dusty annulus of 45~AU in radius that is consistent with the dust morphology observed at shorter wavelengths. 
The VLA observations also reveal a compact source at the center of the disk, possibly due to thermal emission 
from hot dust or ionized gas located within a few AU from the central star. No emission is observed between 
the star and the dusty ring, and, in particular, at the position of the  candidate protoplanet LkCa~15~b. 
By comparing the observations with theoretical models for circumplanetary disk emission, we find that 
if LkCa~15~b is a massive planet ($>5$~M$_J$) accreting at a rate  greater than $10^{-6}$ M$_J$~yr$^{-1}$, then its 
circumplanetary disk is less massive than 0.1~M$_J$, or smaller than 0.4 Hill radii.  Similar constraints are derived 
for any possible circumplanetary disk orbiting within 45 AU from the central star. The mass estimate are uncertain 
by at least one order of magnitude due to the uncertainties on the mass opacity. Future ALMA observations of 
this system might be able to detect circumplanetary disks down to a mass of $5 \times 10^{-4}$~M$_J$ and 
as small as 0.2~AU, providing crucial constraints on the presence of giant planets in the  act of forming 
around this young star.
\end{abstract}

%% Keywords should appear after the \end{abstract} command. The uncommented
%% example has been keyed in ApJ style. See the instructions to authors
%% for the journal to which you are submitting your paper to determine
%% what keyword punctuation is appropriate.


\section{Introduction}
\label{sec:intro}

High angular resolution observations at infrared and millimeter
wavelengths have mapped in great detail nearby young ($<$5 Myr) 
circumstellar disks and revealed ``holes'' 
\citep{Andrews09,Andrews11,Brown08,Brown12,Cieza12a,Hughes09,Isella10a,Isella10b,Isella12,Mayama12,Thalmann10}, 
asymmetric rings \citep[][]{Casassus13,Fukagawa13,Isella13,Perez14,Vandermarel13}, 
and spiral structures \citep{Fukagawa06, Garufi13, Grady13, Hashimoto11, Muto12} in the dust 
spatial distribution.  These features suggest that the observed disks 
are perturbed by low mass companions which remain, to date, elusive.

Detecting giant planets and brown dwarfs orbiting at small separation from young stars with disks
is indeed challenging. The variability of the photospheric lines in young stars and the presence of 
optically thick disks, prevent the use of radial velocities and transit techniques, 
respectively.  Furthermore, direct imaging at optical and infrared wavelengths is feasible only 
if the companions have cleared the surrounding disk regions to expose themselves.  
However, even in this case, current high contrast cameras can image sub-stellar companions only
at angular separations larger  than about 0.1\arcsec, which correspond to orbital 
radii larger than about 15 AU at the distance of nearby star forming regions 
\citep[see, e.g.,][]{Garufi13,Close14}.  Alternatively, detections of sub-stellar companions 
orbiting within dust depleted cavities have been obtained through near-infrared aperture 
masking interferometric observations \citep[][]{Biller12, Huelamo11, Kraus12}, which 
achieve the telescope diffraction limit, e.g., 40 mas for a 10 m telescope at the 
wavelength of 2~$\mu$m. However, 
the sparse nature of aperture masking measurements and the fact they provide information 
only on the closure phase of the Fourier transform of the surface brightness, introduce degeneracies 
in the reconstructions of the source emission.  
%These detections should therefore 
%be considered as tentative until confirmed by other observations.

In this work, we attempt to detect planets in the act 
of forming by observing the millimeter-wave thermal emission from their circumplanetary 
disks. Young giant planets embedded in their primordial nebula are expected 
to be surrounded by circumplanetary disks which regulate the 
angular momentum of the accreting material and provide the raw material to form moons. 
In analogy with circumstellar disks, circumplanetary disks are expected to intercept a large
fraction ($> 20\%$) of the optical and near-infrared radiation from the central planet and 
reemit it at longer wavelengths. 
Attempts to detect circumplanetary disks at millimeter wavelengths have been so far 
inconclusive. \cite{Greaves08} reported the detection of a compact structure in the 1.3 cm 
continuum emission from HL Tau's disk at a radius of 65 AU, which is interpreted as the evidence 
of a circumplanetary disk with a mass of 14 M$_J$. However, the rather low signal-to-noise of 
the detection and the lack of confirmation from observations at shorter wavelengths 
\citep{Carrasco09,Kwon11},  cast doubt on the real nature of the observed structure.

%To our knowledge, this is the first attempt to 
%detect circumplanetary disks around young planets still embedded in their primordial disk. 

The target of our observations is LkCa~15, a 2-5 Myr old  K5 star \citep[L$_\star$= 0.74 L$_\odot$, 
M$_\star =1.0$ M$_\odot$;][]{Simon00,Kenyon95}
located in the Taurus star-forming region at a distance of about 
140 pc \citep[see, e.g.,][]{Loinard07}. The LkCa~15 circumstellar disk has a  
dust-depleted inner region of about 45 AU in radius \citep{Pietu06, Andrews11,Isella12}.
Despite this large cavity in dust, the star is accreting material from the disk at a rate of about $10^{-9}$ M$_\odot$ yr$^{-1}$ \citep{Hartmann98}.
\cite{Kraus12} have reported the discovery of a candidate protoplanet, LkCa~15~b,
through aperture masking observations. The planet candidate is located at a projected 
separation of 70 mas, which corresponds to a physical distance of 16 AU if the planet orbit is 
in the plane of the circumstellar disk. Infrared photometric observations suggest a planet mass 
between 6 and 10 M$_J$ \citep{Kraus12}

Section 2 describes our VLA observations which seek to detect millimeter-wave 
emission from material orbiting around LkCa~15~b, in addition to any circumplanetary disk 
within the dust depleted cavity. A simple radiative transfer model for the 
circumplaneray disk emission is presented in Section 3, and the comparison with the
observations is considered in Section 4. A short discussion of our results and the possibility 
of detecting circumplanetary disks with ALMA is presented in Section 5.

%6 M$_J$ mass protoplanet located inside the continuum
%cavity at an orbital distance of 16AU from the central star. These
%observations suggest that the cavity might be indeed shaped by
%the dynamical interaction with a giant planet. However, Zhu
%et al. (2011) and Dodson-Robinson & Salyk (2011) have argued
%that a cavity 50 AU in radius cannot be explained by a single
%planet, and that additional giant planets or clearing mechanisms
%are required to explain these observations.
%The  
%
%
%
%
%In this letter we present 7~mm high angular resolution VLA observations that target 
%the candidate (proto)planet LkCa15b \citep{Kraus12}.
%The host start, 

\section{Observations}

%\subsection{Observational set up and data reduction}
\label{sec:obs}

\begin{figure*}[!t]
\centering
\includegraphics[angle=0, width=1.0\linewidth]{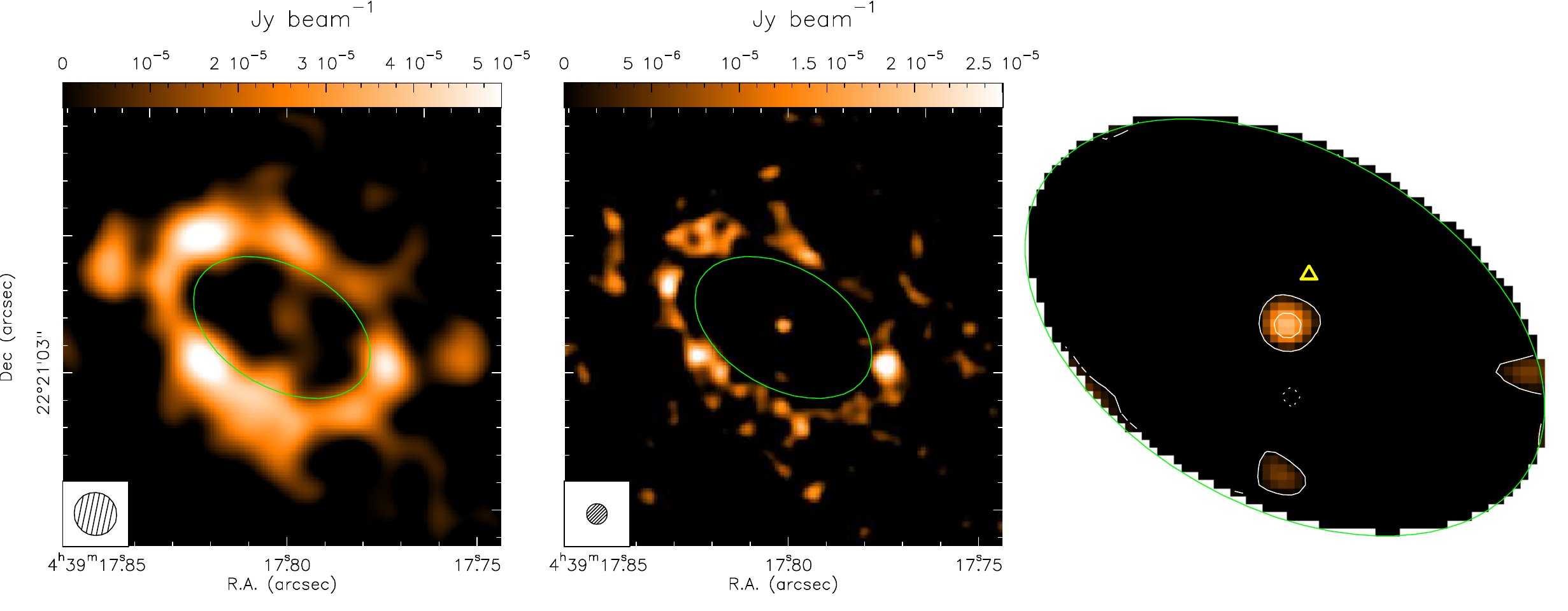}
\caption{
\label{fig:obs} {\it Left:} 1.6\arcsec$\times$1.6\arcsec\ map of the 
LkCa~15's continuum disk emission observed at the wavelength of 7 mm 
obtained by reducing the weights of the complex visibilities measured on the 
longest baselines to increase the sensitivity of the extended structures. 
The rms noise level in the map is 6.1~$\mu$Jy beam$^{-1}$ . The FWHM  
of the synthesized beam is 0.15\arcsec. {\it Center:} map of the 7 mm emission 
obtained by adopting natural weighting of the complex visibilities to maximize the 
angular resolution and the point source sensitivity.  
The rms noise level is 3.6~$\mu$Jy beam$^{-1}$ and the FWHM of the 
synthesized beam is 0.07\arcsec. The green ellipse corresponds to an orbital 
radius of 45 AU and traces the outer edge of the dust depleted cavity as measured from the 
observations at 1.3 mm.  {\it Right:} map of the innermost 45~AU disk region. 
Contours are plotted at 2 and 4$\times$ the noise level.
%The yellow star symbol and the ellipse show the position of LkCa~15 and its uncertainty, 
%respectively, as calculated by applying a proper motion correction of 
%$\delta$RA=-3.9$\pm$3.5~mas and $\delta$Dec=-21.9$\pm$3~mas to 
%the J2000 coordinates of the star \cite[from the UCAC4 catalog][]{Zacharias12}.  
The white triangle shows the expected position of LkCa~15~b \citep{Kraus12} assuming that the star is located at the peak of 
the 7~mm emission.
}
\end{figure*}

LkCa~15 was observed with the VLA in 2012 November, December, and 
2013 January in the A-configuration, which provides baselines between 0.7 and 36 km (project VLA/12B-196).
The total observing time was 68 hours and the time on-source was 21 hours.
The data were obtained using four 2-GHz IFs covering 39 to 47 GHz
(8 GHz total bandwidth) using the Q-band receivers \citep[see][for an overview of the new capabilities of the VLA]{Perley11}. 
The observations used the new 3-bit samplers, and indeed were used to
help commission the hardware through the EVLA Commissioning Staff
Observing (ECSO) program.  The data were calibrated using the CASA
data reduction package \citep[e.g.,][]{McMullin07}, and a version
of the VLA Calibration Pipeline scripts (see
\url{https://science.nrao.edu/facilities/vla/data-processing/pipeline})
modified to average data across each 2-GHz IF to enable the use of
a nearby, weak source ICRF J042655.7+232739 (0.2~Jy) as the complex gain
calibrator.  3C147 was observed as the primary flux density calibrator
\citep{Perley13}, and 3C84 was observed as the bandpass
calibrator.  The absolute uncertainty in the overall flux density
scale is estimated to be 10\%.

%\begin{figure}
%\centering
%\includegraphics[angle=0, width=0.9\columnwidth]{FIGURE/cont_ABCE_uvamp.eps}
%\caption{\label{fig:corr} The top and bottom panel show the real and imaginary part of the 
%correlated flux respectively as a function of the baseline length $B_{uv}$. The flux has been circularly averaged 
%on 75 m wide radial bins after shifting the phase center of the observations at the position corresponding to the 
%center of symmetry of the emission }
%\end{figure}

%\subsection{Results}

Figure 1 presents the maps of the 7~mm continuum emission observed 
toward LkCa~15 obtained using the CLEAN task in CASA with two different 
weighting schemes of the complexed visibilities. The map on the left was obtained by 
reducing the weights of the longest baselines (tapering) to increase the sensitivity on 
the extended structures, while the central and right maps were obtained by  
adopting natural weighting to maximize the angular resolution and point source sensitivity.
In this latter case, the observations achieve an angular resolution of 
70~mas, or 9 AU at the distance of the star, and a rms noise of 3.6~$\mu$Jy beam$^{-1}$. 
The total flux density measured within the shown region is 0.52$\pm$0.06 mJy and its 
consistent with the value of 0.44$\pm$0.17 mJy measured by \cite{Rodmann06} using the more 
compact D configuration. This suggest that although our observations are insensitive to 
spatial scales larger than 2\arcsec, they recover most of dust emission. The spectral index of 
the millimeter wave emission ($F_\nu \propto \nu^\alpha$) measured between 0.87~mm 
and 7~mm is 3.2$\pm$0.2. 

The observations  reveal that most of the emission arises from an elliptical annulus characterized 
by a major axis of 0.8\arcsec, a minor axis of  0.5\arcsec, and a position angle of about 60\arcdeg, 
as measured from north toward east.  The size and orientation of this annulus match the 
geometry of the dust emission observed at 0.87~mm and 1.3~mm \citep{Pietu06,Andrews11,Isella12}, 
and confirm that LkCa~15's disk is depleted of dust within a radius of about 45 AU, as indicated 
by the green ellipse. In the tapered map, the dust emission is characterized by a clumpy 
structure on spatial scales comparable with the resolution of the observations, i.e., 20 AU. 
More precisely, the intensity varies azimuthally along the ring from 18 to 50 $\mu$Jy beam$^{-1}$. 
The significance of these structures will be discussed in a forthcoming paper that study the morphology 
of the dusty ring as revealed by the multi-wavelength continuum data.

The 7 mm map reveals a point source with a flux density of 16.6 $\mu$Jy beam$^{-1}$ 
located at the center of the dusty ring.  Because its coordinates are consistent with 
the proper motion corrected coordinates of LkCa~15, we will assume in the following 
that this compact source pinpoints the position of the central star.  
The observations at shorter wavelengths do not detect any unresolved emission 
at the center of the disk. More precisely, \cite{Andrews12} find that the disk regions within 
the dust depleted cavity might contribute up to about 5 mJy to the observed 870~$\mu$m emission, while 
the 1.3 mm complex visibilities measured on baselines longer than 500 k$\lambda$ with CARMA set 
a 3$\sigma$ upper limit  of 1.5 mJy beam$^{-1}$ for compact emission at the center of the disk \citep{Isella12}. 
These measurements imply that the compact emission observed at 7 mm has a spectral index $\alpha <2.7$, 
which is smaller than spectral index derived for the spatially integrated disk emission.

The compact emission is not consistent with the radiation from the stellar photosphere, which has 
an expected 7~mm flux of 0.06 $\mu$Jy.
Alternatively, we suggest that it might arise either from a narrow ring of 
millimeter size grains orbiting within a few AUs from the central star, or from 
ionized gas in the vicinity of the star \citep[see, e.g.,][]{Panagia74}.  In the former 
case, the measured flux would suggest the presence of about 3~Earth masses of 
small dust grains, as calculated by assuming optically thin emission, a dust temperature of 200 K, 
and a dust opacity at 7~mm of  $0.2$ cm$^{2}$ g$^{-1}$.
Investigating the nature of the compact emission requires high angular resolution 
observations at different wavelengths in order to measure its spectral index.

With the exception of the central source, we do not detect any emission above 3$\times$ the noise 
level inside the dust depleted cavity, and,  in particular, at the expected position of 
LkCa~15~b \citep{Kraus12}. The constraints that this non detection sets on the structure of 
circumplanetary disks orbiting within the dust depleted cavity are discussed in the 
following sections.

%\section{Analysis}

\section{Circumplanetary disk model}
\label{sec:model}

%\subsection{Disk surface density and mass}
We adopt a circumplanetary disk model characterized by a power-law 
surface density $\sigma(r) \propto r^{-p}$ that extends from the planet radius 
out to some outer radius $r_d$, at which the material is  captured in orbit 
around the central star.  Numerical simulations suggest that the 
truncation of a circumplanetary disk by the stellar gravitational field 
might occur  between 0.1 $r_H$ and 0.7 $r_H$ \citep[see, e.g.,][]{Canup02,DAngelo02,Szulagyi13,Ward10}, where $r_H = R_p  \sqrt[3]{M_p / 3M_\star}$
is the planet Hill radius, $R_p$ is the planet orbital radius, $M_p$ is the planet mass, 
and $M_\star$ is the mass of the central star. 
%In our model, $r_d$ is a free parameter
%which is allowed to vary between 0.01 and 1 $r_H$. 
% In the case of LkCa15b, the planet mass 
%is loosely constrained between  5 and 10~M$_J$, while the orbital radius is 
%16 AU \citep{Kraus12}. For $M_\star = 1 M_\sun$ (ref), the LkCa15b's Hill radius 
%should therefore be between 1.9 and 2.4 AU.  
As in the analytical models of \cite{Canup02}, we define a nominal disk model with $p=3/4$. 
The effects that this choice has on the results are discussed in Section~\ref{sec:res}.
For an inner disk radius $r_{in} \ll r_d$, the disk surface density can be integrated to 
give the total disk mass
\begin{equation}
M_d = 2\times10^{-2} \left( \frac{\sigma_{1\textrm{AU}} }{36 \, \textrm{g cm}^{-2}}   \right)  \left( \frac{r_d}{\textrm{1 AU}} \right) ^{5/4} M_J,
\end{equation}
where $\sigma_{1\textrm{AU}}$ is the surface density at 1 AU, and 
the normalization constant corresponds to the minimum mass with solar composition
required to form the four largest moons of Jupiter, i.e., the 
Galilean satellites \citep[][]{Pollack84}.

%``minimum mass subnebula" 
%({\it mmsn} hereafter) , i.e., the 

%The masses of the  suggest that they formed in a 
%gas-rich circumplanetary disk containing about 2\% of Jupiter's mass. 
%In analogy with the ``minimum mass solar nebula'', this value is generally 
%referred as the `t is used as normalization
%value in Equation 1. The circumplanetary disk mass and outer radius are
%kept as free parameters in our model and we will explore values 
%between $10^{-6}$ and $10^{-2}$ M$_\sun$, and 0.2 AU and 1.7 AU, 
%respectively. 
%
%\subsection{Disk temperature and emission}

The temperature of the circumplanetary disk at the distance 
$r$ from the planet is expressed as 
\begin{equation}
T_d^4(r) = T_{irr,\star}^4(R) + T_{irr,p}^4(r) + T_{acc}^4(r), 
\end{equation} 
where $T_{irr,\star}$ corresponds to the irradiation from the 
central star at the orbital radius $R$, $T_{irr,p}$ corresponds 
to the irradiation from the planet itself, and $T_{acc}$ is the contribution 
from the viscous dissipation within the circumplanetary disk.
The disk temperature  is calculated by extending the ``two-layer" radiative transfer 
model developed by \cite{Chiang97} in the context of circumstellar disks 
to the case of circumplanetary disks.  The model assumes that the disk 
is in hydrostatic equilibrium between the gas pressure and the planet gravity, 
and the disk temperature is calculated by iterating on the vertical structure.  
The implementation of the model is the same as 
in \cite{Dullemond01} \citep[see also][]{Isella09,Isella10a,Isella12} where the stellar 
effective temperature, radius, and luminosity are replaced by the values 
proper for the planet, and the circumstellar disk properties are 
replaced by the values corresponding to the circumplanetary disk. 

The heating contribution from the planet depends on the planet
effective temperature and radius, i.e., $T_{irr,p}^4 \propto T_p^4 (r_p/r)^2$. 
We adopt \cite{Spiegel12} planetary models, which predict that the 
temperature and radius of a young planet depend on the formation process. 
In brief, planets formed by gravitational instability are expected to be hotter 
and more luminous than planets formed through the core accretion process. 
For the sake of simplicity, these two models are identified as ``Hot start" 
and  ``Cold start", respectively. For example,  a 2 Myr old 10~M$_J$ planet 
is predicted to have $T_p \sim 2500$~K and a $r_p \sim 1.7$ $R_J$ 
in the ``Hot start'' scenario, and $T_p \sim 700$~K and $r_p = 1.2$ $R_J$
for the ``Cold start'' case. 

We assume that the circumplanetary disk is directly illuminated by the central star 
and calculate the stellar heating using the ``two-layer" disk model where, this time, 
the source of radiation is the central star LkCa~15. Following the result of \cite{Isella12}, 
we assume  $T_{irr,\star}(R) = 40\textrm{K} \times (R/16\textrm{AU})^{-1/2}$.

Finally, the energy released per unit area in an annulus at distance $r$ from 
the planet by accreting material can be expressed as  \citep{Dalessio98}
\begin{equation}
\label{eq:Tacc}
T_{acc}^4(r) = \frac{3G M_p M_{acc}}{8\pi \sigma_{SB} r^3}\left[ 1- \left( \frac{r_p}{r} \right)^{1/2} \right],
\end{equation}
where  $M_{acc}$ the mass accretion rate onto the planet and $\sigma_{SB}$ is the Stefan--Boltzmann 
constant.

%  is a free parameter 
%that is allows to vary between $10^{-5}$ M$_\sun$ yr$^{-1}$ and $10^{-9}$  M$_\sun$ yr$^{-1}$.
%
%Note that the mass accretion rate on LkCa15 is $2 \times 10^{-9}$ M$_\sun$ yr$^{-1}$.

\begin{figure}[t]
\centering
\includegraphics[angle=0, width=0.9\linewidth, bb=70 0 290 230, clip=True]{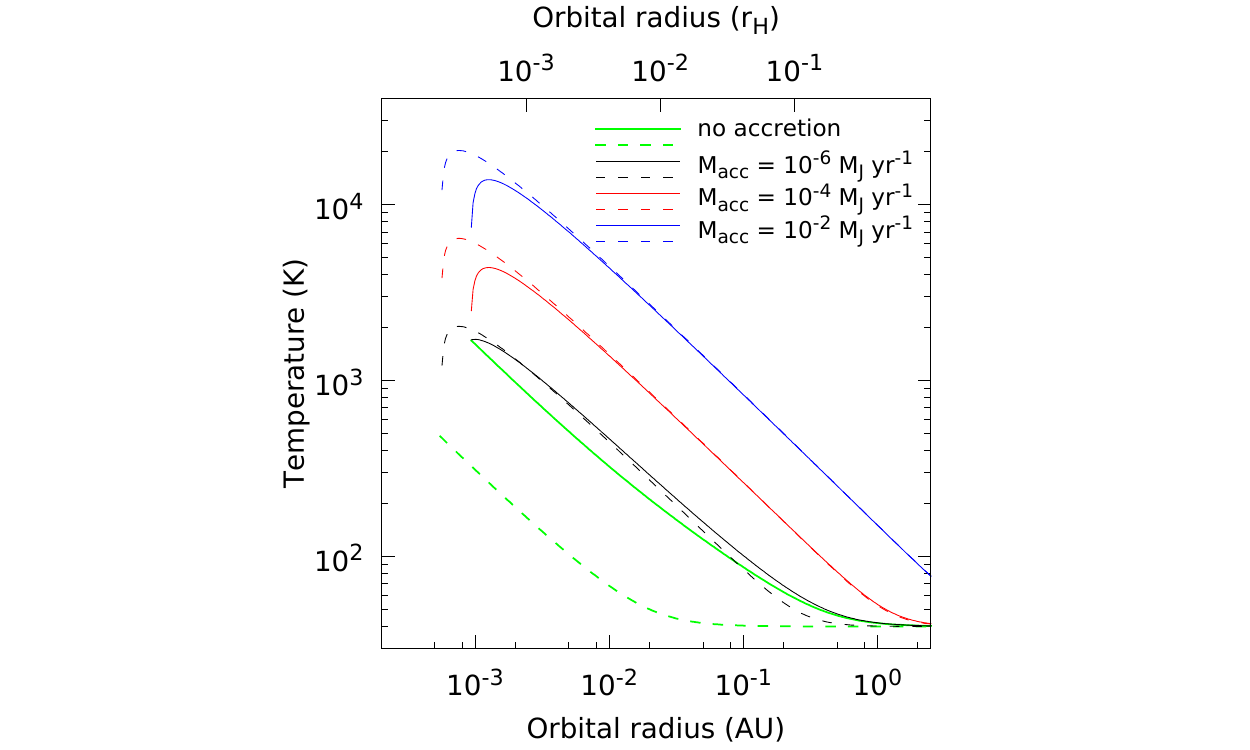}
\caption{\label{fig:disktemp} Interior temperature of a circumplanetary disk as 
a function of the distance from a 10~$M_J$ planet orbiting at 16 AU from LkCa~15 ($M_\star = 1 M_\odot$, $L_\star = 0.74 L_\odot$). 
The solid and dashed curves correspond to 
planet effective temperatures of 2500 K and 700 K, respectively. The different colors indicate 
models characterized by different mass accretion rates. From top to bottom: blue, $M_{acc} = 10^{-2}$ $M_J$ yr$^{-1}$;
red, $M_{acc} = 10^{-4}$ $M_J$ yr$^{-1}$; black, $M_{acc} = 10^{-6}$ $M_J$ yr$^{-1}$; 
green, no accretion. The viscous heating dominates over the irradiation from the planet 
and from the star for $M_{acc} \geq 10^{-6}$ $M_J$ yr$^{-1}$. }
\end{figure}

We show in Figure~\ref{fig:disktemp} the circumplanetary disk temperature 
calculated in the case of a 10 M$_J$ planet characterized by effective 
temperatures of  2500 K and 700 K, which are representative of the 
``Hot start" and ``Cold start" models. The disk temperature is calculated  
for mass accretion rates of  0, $10^{-6}$, $10^{-4}$, 
and $10^{-2}$ M$_J$ yr$^{-1}$. These latter three values correspond to the mean 
accretion rate required to build up a 10 M$_J$ planet in 10$^7$, $10^5$, and $10^3$ yr 
and might be representative of different accretion phases during the evolution of the 
planet and its parent disk.
We find that for $M_{acc} \geq 10^{-6}$ M$_J$ yr$^{-1}$ the disk temperature is 
dominated by the accretion contribution expressed by Equation~\ref{eq:Tacc}, 
and it is in first approximation independent from the assumed planet model. 
At $M_{acc} \geq 10^{-6}$ M$_J$ yr$^{-1}$ the disk temperature can 
exceed the sublimation temperature of silicates grain which is around 1500~K. 
In this case, and since the disk opacity at millimeter wavelengths is dominated 
by the dust,  we adopt a disk inner radius $r_{in}$ consistent with the dust 
sublimation radius. 

%\subsection{Disk continuum emission at millimeter-wavelengths}
 
The disk emission is then calculated as 
\begin{multline}
\label{eq:flux}
F_\textrm{7mm} = 2\pi \cos{i} \int_{r_{in}}^{r_d} \left\{1-\exp \left[ \frac{-\sigma(r) 
\kappa_\textrm{7mm}}{\cos{i}} \right] \right\} \times \\ \times B_\nu[T_d(r)] \, \frac{r}{d^2} \, dr,
\end{multline}
where the mass opacity  $k_\textrm{7mm} = 0.002$~cm$^{2}$ g$^{-1}$ is 
calculated by assuming a dust-to-gas ratio of 0.01 and a dust composition and 
grain size distribution as in \cite{Isella12}. We discuss the effects of relaxing this 
assumption in Section~\ref{sec:disc}.

\section{Results}
\label{sec:res}

\begin{figure*}[]
\centering
\includegraphics[angle=0, width=0.33\linewidth, bb=80 0 260 220, clip=True]{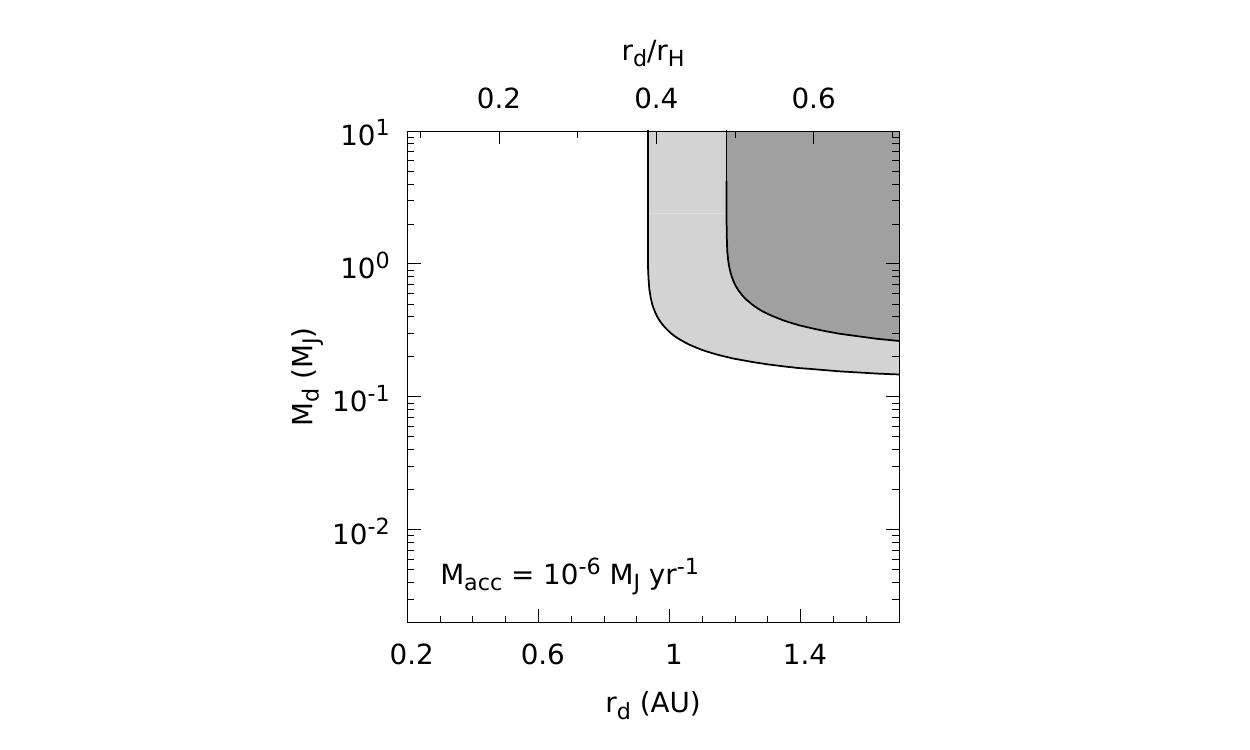}
\includegraphics[angle=0, width=0.33\linewidth, bb=80 0 260 220, clip=True]{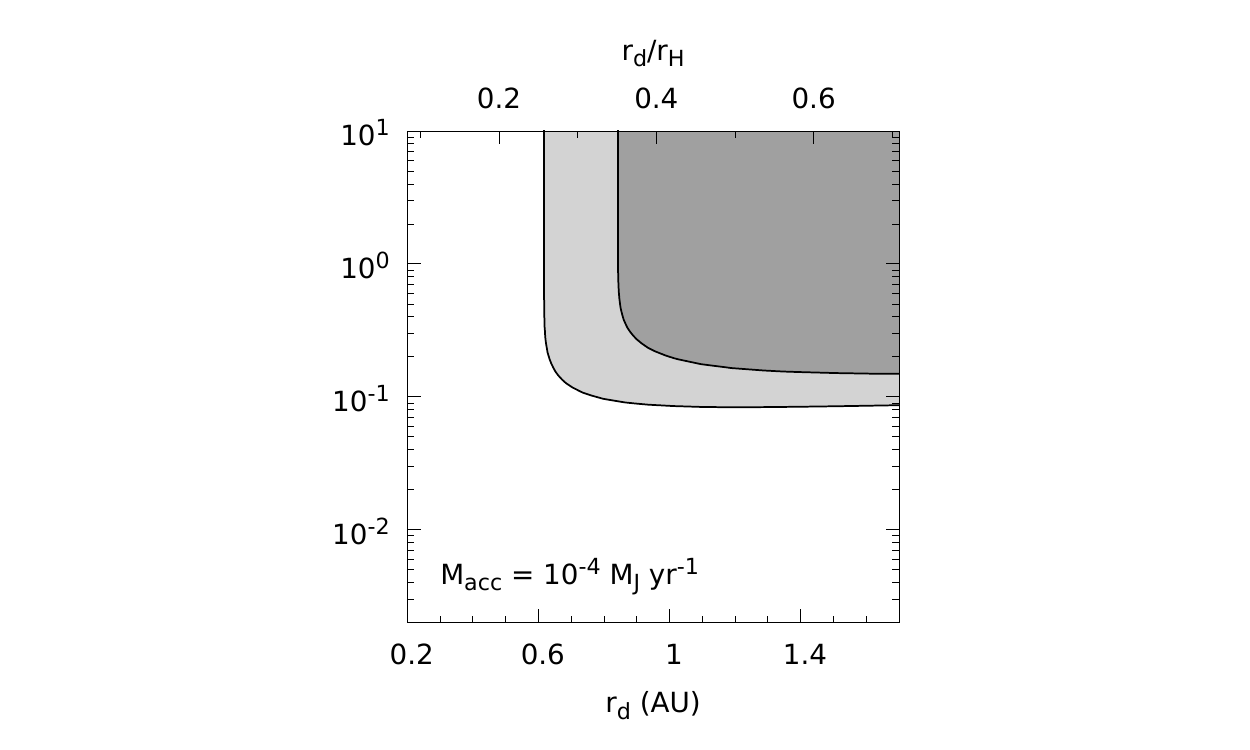}
\includegraphics[angle=0, width=0.33\linewidth, bb=80 0 260 220, clip=True]{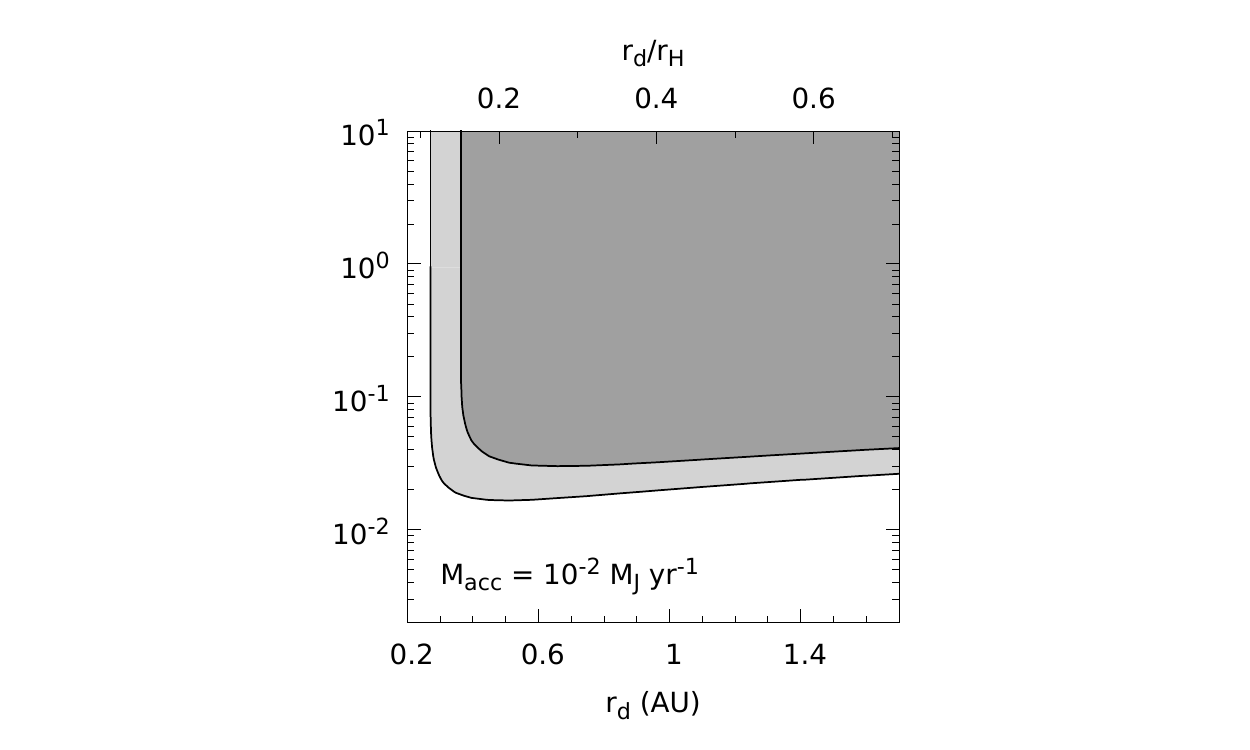}
\caption{\label{fig:mod} The light and dark gray regions show the disk models characterized by an integrated 
continuum flux at 7 mm larger than 2$\times$ and 3$\times$ the noise level of our observations, i.e., 3.6 $\mu$Jy beam$^{-1}$.
Since the observations do not reveal any emission at the position of LkCa~15~b, these circumplanetary disk 
models are ruled out at  more than 95.4\%  and 99.7\% confidence level, respectively.  
The disk emission has been computed for a planet characterized by M$_p = 10$~M$_J$, $T_p = 2500$~K, 
and $r_p = 1.7 R_J$, and for mass accretion rates of $10^{-6}$ (left), $10^{-4}$ (central), and $10^{-2}$~M$_J$ yr$^{-1}$ (right).}
\end{figure*}

We adopt the model discussed above to calculate the emission of a possible disk orbiting 
around the candidate planet LkCa~15~b, and then we generalize the results to the case of 
 any circumplanetary disk within the LkCa~15's disk dust depleted cavity.

In the case of LkCa~15~b, we assume a planet mass $M_p=10M_J$, 
and an orbital radius $R_p = 16$~AU \citep{Kraus12}.  We adopt a planet temperature 
$T_p = 2500$~K and a planet radius  $r_p=1.7$~R$_J$, that correspond to the 
``Hot start" model for a 2 Myr old planet of 10 M$_J$ \citep{Spiegel12}. 
The circumplanetary disk model has three free parameters: the disk outer radius, the disk mass, 
and the mass accretion rate. We assume a circumplanetary disk inclination of 45\arcdeg\, i.e., the same inclination 
of the circumstellar disk. We calculate the disk emission for disk outer radii 
between 0.2 AU and 1.7 AU, corresponding to 0.1~$r_H$ and 0.7~$r_H$, respectively. 
Disk masses vary between  $10^{-3}$~M$_p$ and  M$_p$, and the mass accretion rate 
ranges between $10^{-6}$ and $10^{-2}$ M$_J$ yr$^{-1}$. 
In the next section we argue that $M_{acc} = 10^{-6}$ M$_J$ yr$^{-1}$ might be the 
 most appropriate value for the LkCa~15 system.

\begin{figure}[t]
\centering
\includegraphics[angle=0, width=0.7\linewidth, bb=80 31 260 200, clip=True]{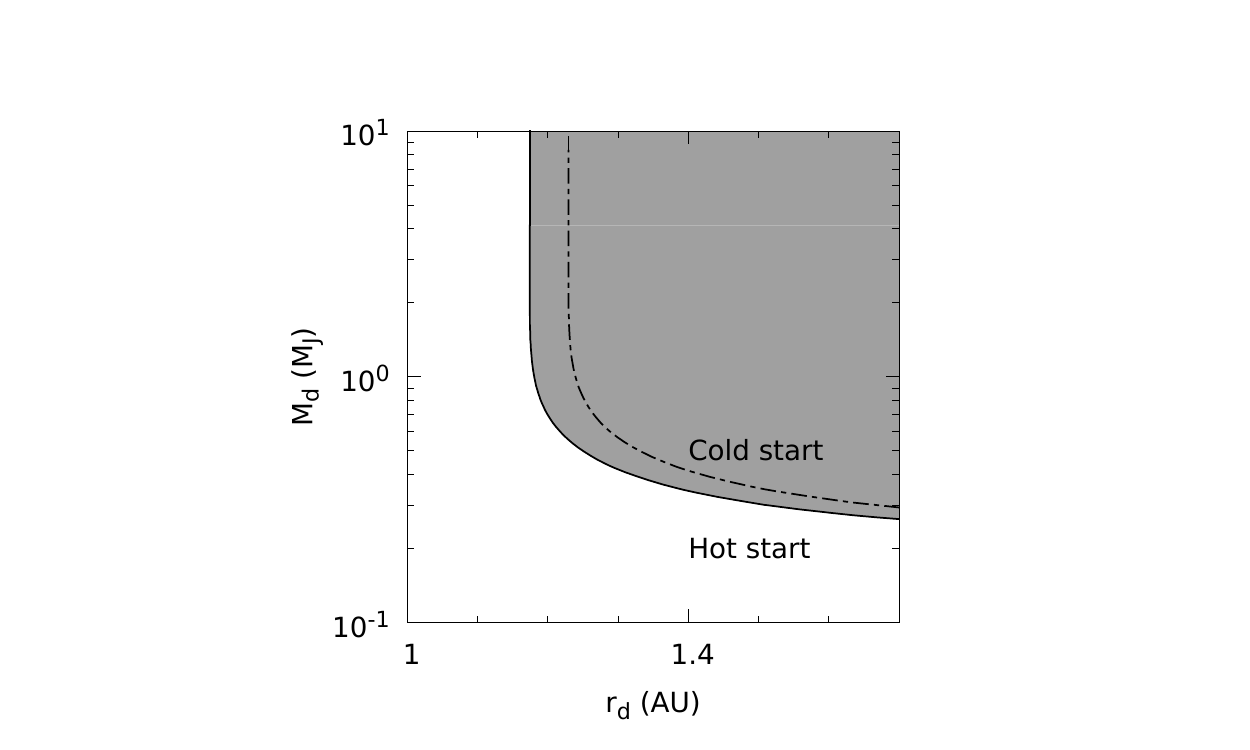}
\includegraphics[angle=0, width=0.7\linewidth, bb=80 0 260 190, clip=True]{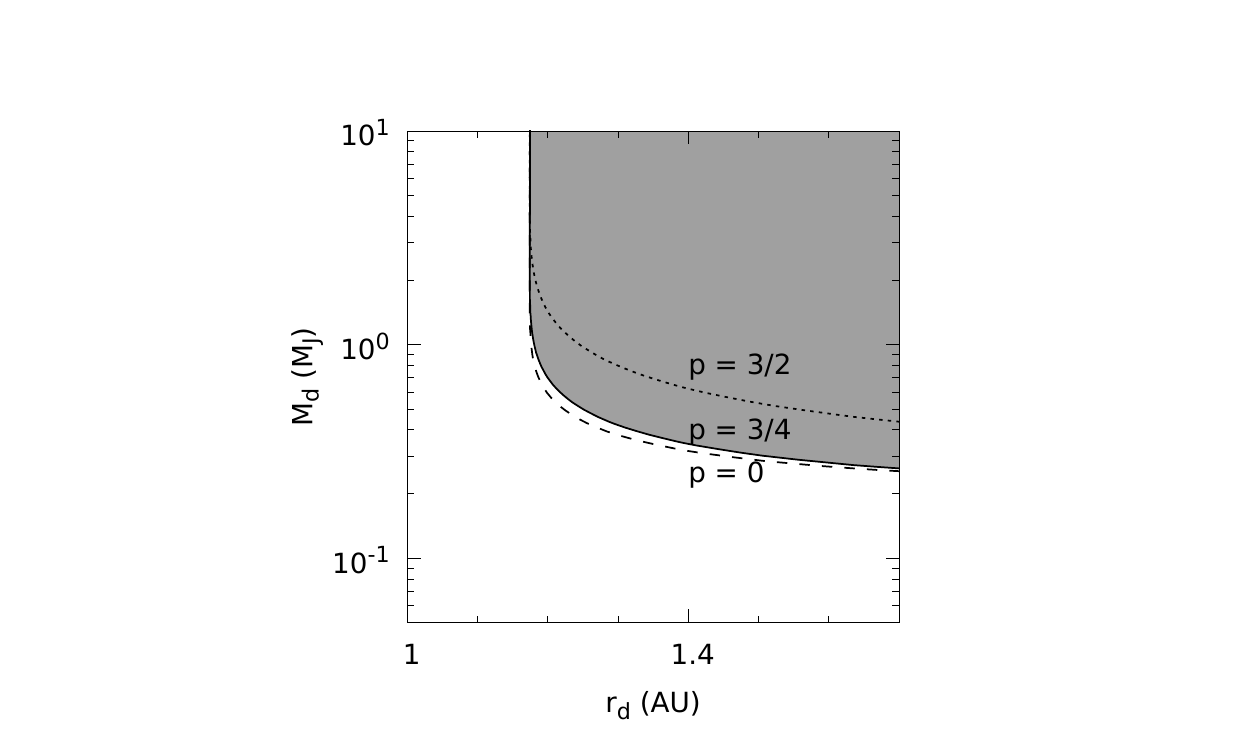}
\caption{\label{fig:mod_comp} Dependence of the constraints on the LkCa~15~b circumplanetary disk structure on the 
planet model (top) and on the slope of the surface density (bottom). As in Figure~3, the dark gray region shows 
model with integrated fluxes at 7 mm larger than 3$\times$ the noise level of our observations. The top panel shows
the comparison between the nominal ``Hot start" model for LkCa~15~b (age = 2~Myr, M$_p = 10$~M$_J$, $T_p = 2500$~K, 
and $r_p = 1.7 R_J$) and a ``Cold start" model (age = 2 Myr, M$_p = 10$~M$_J$, $T_p = 700$~K, 
and $r_p = 1.2 R_J$) for $M_{acc}= 10^{-6}$ M$_J$ yr$^{-1}$. The bottom panel compares the nominal disk 
model ($p=3/4$) to the case of a flatter ($p=0$) and steeper ($p=3/2$) surface density profile.}
\end{figure}

Because the circumplanetary disk radius is much smaller that the angular resolution 
of the our observations, we compare the theoretical 7~mm integrated disk flux 
to the point source sensitivity of the observations, i.e, 3.6~$\mu$Jy beam$^{-1}$. 
The gray regions in Figure~\ref{fig:mod} show the disk models with 7~mm fluxes larger 
than 2$\times$ and 3$\times$ the noise level of our observations. Given the non detection 
at the position of LkCa~15~b, these models are ruled out with a probability higher than 95.4\% 
and 99.7\%, respectively. 
In general, the non detection of LkCa~15~b disk rules out the presence of a 
large and massive circumplanetary disk. If the 7~mm disk emission is optically thin, the observations 
set an upper limit of the disk mass which is in first approximation independent of the 
disk radius and scales with the mass accretion rate as $M_{acc}^{-1/4}$.
The optically thin limit is shown by the almost horizontal lower margins of the gray regions
in Figure~\ref{fig:mod}. By contrast, if the 7~mm disk emission is optically thick, 
the observations set an upper limit on the disk outer radius that is independent on 
the disk mass and scales as $M_{acc}^{-1/8}$. A variation of four orders of 
magnitude in $M_{acc}$ therefore corresponds to a variation of a factor 3 
in the upper limit for the outer disk radius for optically thick disks and of a factor 10 
in upper limit on the disk mass in the optically thin case.

Figure~\ref{fig:mod_comp} shows the constraints on the disk mass and radius 
for $M_{acc} = 10^{-6}$ M$_J$ yr$^{-1}$ in the case of a colder planet (top panel) and for 
different slopes of the disk surface density (bottom panel).
Colder planets lead to a fainter millimeter disk emission, but as far as the 
disk temperature is dominated by the accretion heating, the planet temperature 
has a minor effect on the overall disk emission. We find that the disk flux is 
anti-correlated with the slope of the surface density, and, consequently, the
upper limit on the disk mass diminishes for lower values of $p$.  
This is due to the fact that in the range of disk masses and radii shown in the figure, 
a large fraction of the 7~mm comes from the innermost optically thick regions of the disk.
In this case, a steeper surface density implies that there is less mass in the outer 
optically thin region, which causes a decrement in the flux.
 
\begin{figure*}[!t]
\centering
\includegraphics[angle=0, width=0.33\linewidth, bb=80 0 260 220, clip=True]{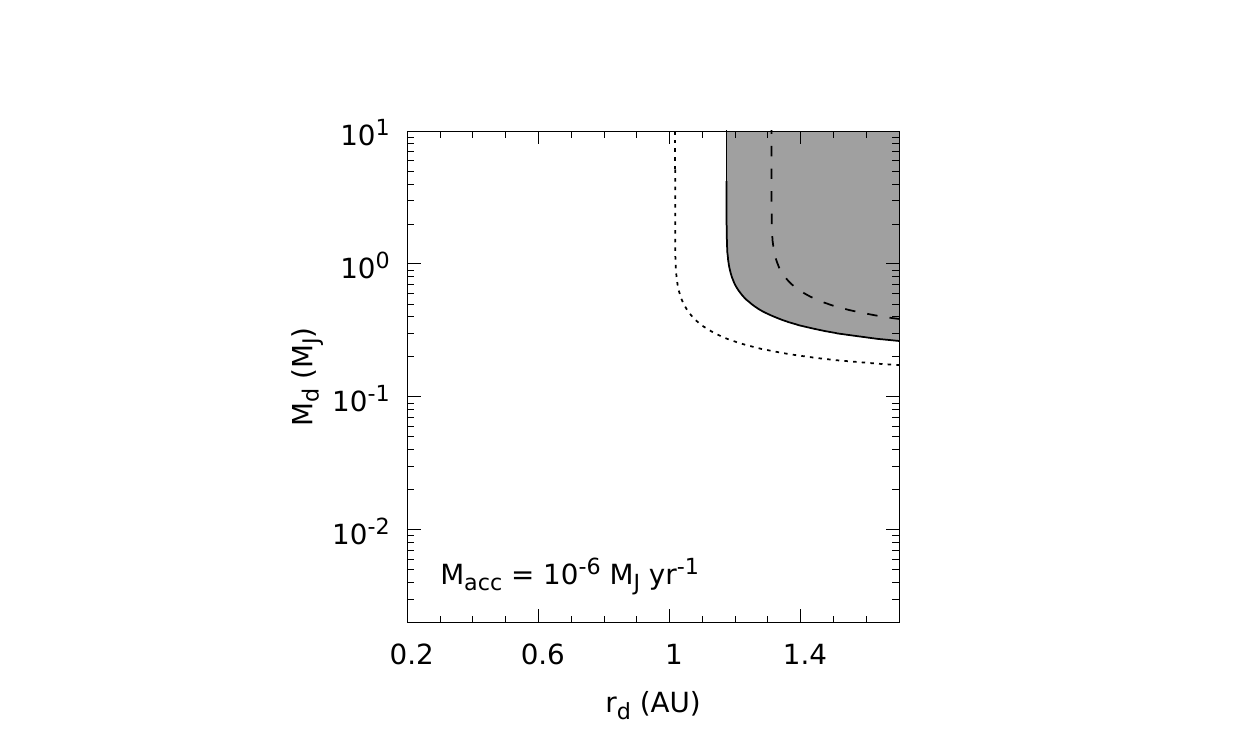}
\includegraphics[angle=0, width=0.33\linewidth, bb=80 0 260 220, clip=True]{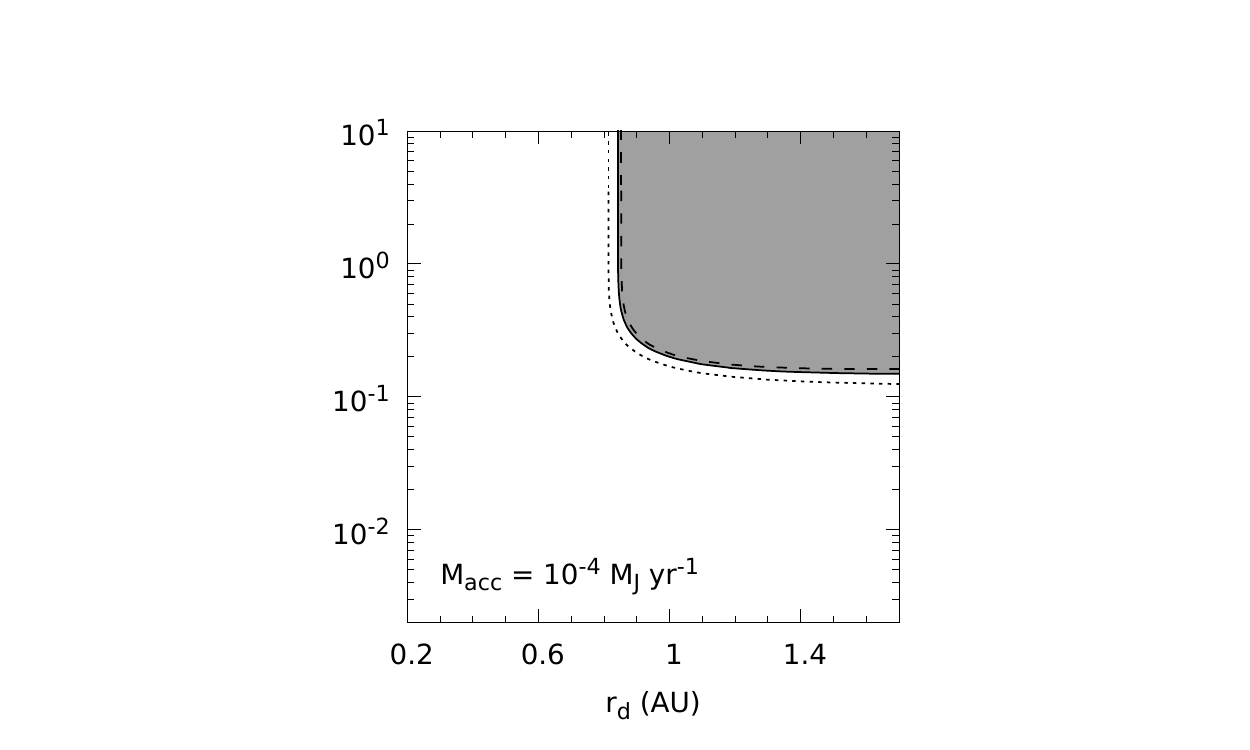}
\includegraphics[angle=0, width=0.33\linewidth, bb=80 0 260 220, clip=True]{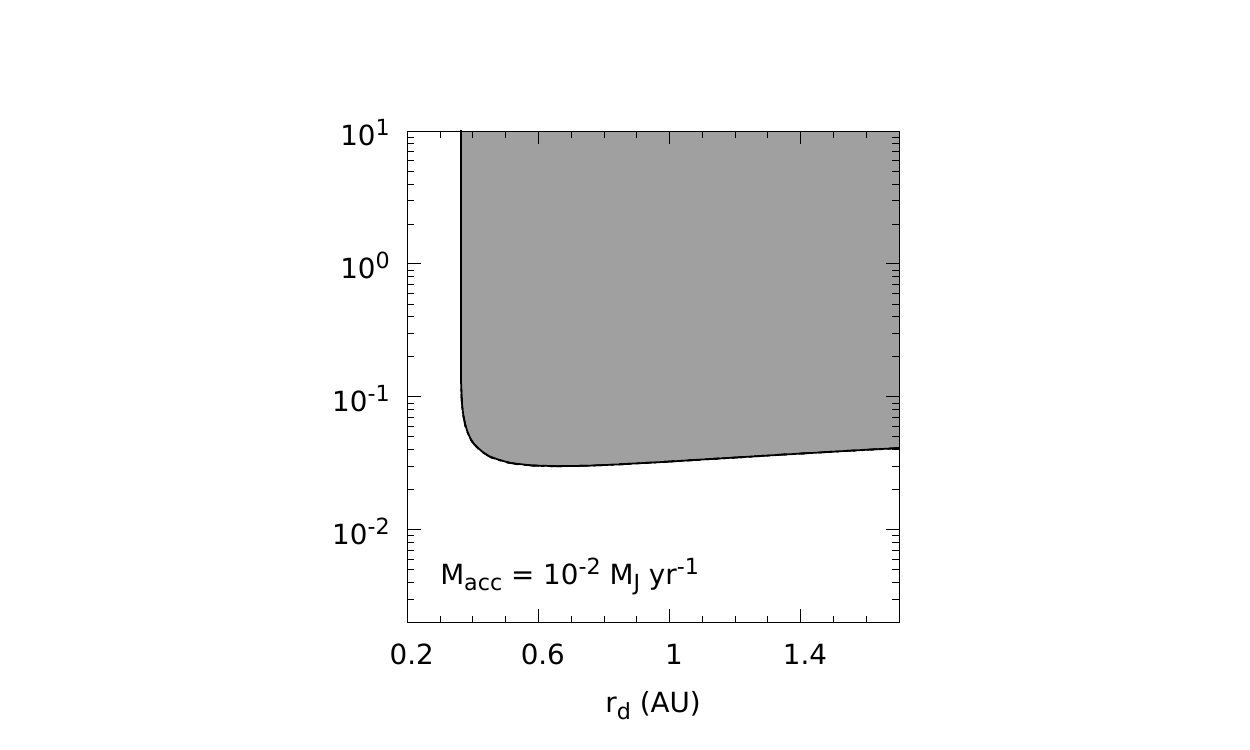}
\caption{\label{fig:comp_radius}  Dependence of the constraints on the circumplanetary disk structure on the 
planet orbital radius.   As in Figure~3, the dark gray region shows model with integrated fluxes at 7 mm larger 
than 3$\times$ the noise level of our observations in the case of circumstellar disk orbiting at 16 AU from the central star. 
The dotted and dashed curves indicate circumstellar disks with 7~mm fluxes equal to  3$\times$ the noise level 
orbiting at 8 AU  and 32 AU from the central star, respectively.}
\end{figure*}

The results obtained for the LkCa~15~b disk can be generalized to giant planets with 
different masses and orbital radii  by considering that (1) for $M_{acc} > 10^{-6}$ 
 M$_J$ yr$^{-1}$ the disk emission does not depend on the planet luminosity and 
 (2) the heating contribution from the central star scales as $R^{-2}$. Circumplanetary 
 disks orbiting closer to the star are therefore expected to be hotter and more luminous 
 at millimeter wavelength, while disks orbiting further out should be colder and fainter. 
 The overall effect on the constraints on the disk properties is shown in Figure~\ref{fig:comp_radius}, 
 which compare the nominal model for LkCa~15~b, to that of 
similar disks orbiting at 8 and 32 AU, i.e., half and twice the LkCa~15~b orbital radius 
respectively. The effect of the orbital radius on the disk emission are 
larger for low mass accretion rate values and become negligible 
at $M_{acc} = 10^{-2}$ M$_J$ yr$^{-1}$ since the disk temperature 
is totally dominated by the accretion heating.

%\section{Predictions for ALMA observations}

\section{Discussion and Conclusions}
\label{sec:disc}
LkCa~15 is surrounded by a gas rich circumstellar disk that accretes onto 
the central star at a rate of $1.3 \times 10^{-9}$ M$_{\sun}$ yr$^{-1}$, as 
measured by calculating the accretion luminosity from the ultra-violet excess emission 
over the stellar photosphere \citep{Hartmann98}. 
Hydrodynamic simulations of giant planets embedded in gaseous 
rich circumstellar disks suggest that the amount of material accreting 
from the circumstellar disk onto a circumplanetary disk might be comparable 
to the mass accretion rate onto the central star. If circumplanetary disks have 
a viscosity similar to that of the circumstellar disks, then the mass accretion rate 
onto the planet should also be comparable to the mass accretion rate onto 
the central star \citep{Zhu11,Szulagyi13}.  This suggests that
giant planets orbiting inside the dust depleted cavity might be accreting, in average,  
at a rate of  $10^{-9}$ M$_{\sun}$ yr$^{-1}$, or  $10^{-6}$ M$_J$ yr$^{-1}$.
This value corresponds to the minimum mean mass accretion rate 
required to form a giant planet such as LkCa~15~b faster than
 the  average  disk dispersal time scale derived from infrared observations \citep{Hernandez08}. 
 Higher accretion rates are predicted during 
 the initial phase of  planet formation \citep{Ward10,Shabram13}, while 
 lower values are possible if circumplanetary disks are less viscous than 
 the circumstellar disk. However, this would imply that any giant 
 planet orbiting LkCa~15 has, in practice, accreted the majority of its final 
 mass despite the young age of the system.

The non detection of millimeter emission at the position of the 
candidate young planet LkCa~15~b, and, in general, within the dust depleted 
cavity in LkCa~15 circumstellar disk, sets upper limits on the mass and radius 
of possible circumplanetary disks. For $M_{acc} \geq10^{-6}$ M$_J$ yr$^{-1}$, we find that 
the disk temperature is dominated by the viscous heating released by the accreting material 
and the 7~mm continuum disk emission is in first approximation independent  of the luminosity 
of the central planet. 
Under this condition, our observations exclude the presence of disks more massive 
than about 0.1 $M_J$  and larger than about 1~AU, or 0.4 $r_H$ for a 10 $M_J$ planet orbiting at 16 
AU from the central star. 
Higher mass accretion rates would imply lower disk masses and radii, but, as discussed in the previous section, an 
increase of four orders of magnitude in the mass accretion rate would lower the constraints on
disk mass only by a factor of 10 and the disk radius by a factor of 3. 
If the dust depleted cavity observed in the LkCa~15 disk originates from the dynamical clearing 
operated by brown dwarfs or massive planets (e.g.,$>10$ M$_J$), our VLA observations suggest 
that the mass of their circumstellar disks should be less than a few percent of the planetary mass, 
unless they have very small radii. 

The upper limits for the circumplanetary disk mass are derived adopting the 
dust model employed in the study of the LkCa~15 circumstellar disk emission, which  
assumes a grain composition as in \cite{Pollack94}, a grain size distribution $n(a)\propto a^{-3.5}$ 
ranging between 0.5~$\mu$m and 0.5~mm, and a gas-to-dust ratio of 100 \citep[][]{Isella12}.
The corresponding mass opacity at 7~mm is $2\times10^{-3}$ cm$^2$ g$^{-1}$.
There are however at least two (competing) physical processes that might cause 
the mass opacity of a circumplanetary disk to differ from that of the parent circumstellar disk. 
The first process is the filtration, or trapping, of large dust grains at the outer  
edge of the cavity cleared by the gravitational interaction with massive planets. 
More precisely, \cite{Zhu11} find that only dust grain smaller than 10-100~$\mu$m might 
be able to filtrate from the outer disk into the dust depleted 
cavity. In this scenario, the material accreting onto circumplanetary disks would be 
depleted by large dust grains and have a gas-to-dust ratio higher than the circumstellar 
disk material. For example,  if all the grains larger than 10~$\mu$m 
are trapped in the outer disk at $R>45$ AU, then the material accreting onto the circumplanetary 
disk would have a gas-to-dust ratio of 1000, i.e., 10 times larger than the material 
in the outer disk.  The resulting mass opacity at 7 mm  calculated accounting
for both the higher gas-to-dust ratio and for the reduced maximum grain size would be  
$1.6\times10^{-4}$ cm$^2$ g$^{-1}$.
However, the dust grains in a circumplanetary disk grains are expected to interact and 
grow in size similarly to what happens in the innermost regions of circumstellar disks.
Actually, grain growth in circumplanetary disks is a key ingredient in current models 
for the formation of  the Galilean satellites \citep[see, e.g.,][]{Canup09}.
In this case, if the maximum grain size grows, for example, from 10~$\mu$m to 1 cm, then the 7 mm 
mass opacity would increase to $1.8\times10^{-3}$ cm$^2$ g$^{-1}$ (now
assuming a gas-to-dust ratio of 1000). Given the uncertainties on mass opacity, the 
constraints in the circumplanetary disk mass set by our observations should therefore be cosidered 
uncertain by at least one order of magnitude.

\begin{figure}[!t]
\centering
\includegraphics[angle=0, width=\linewidth, bb=80 0 260 220, clip=True]{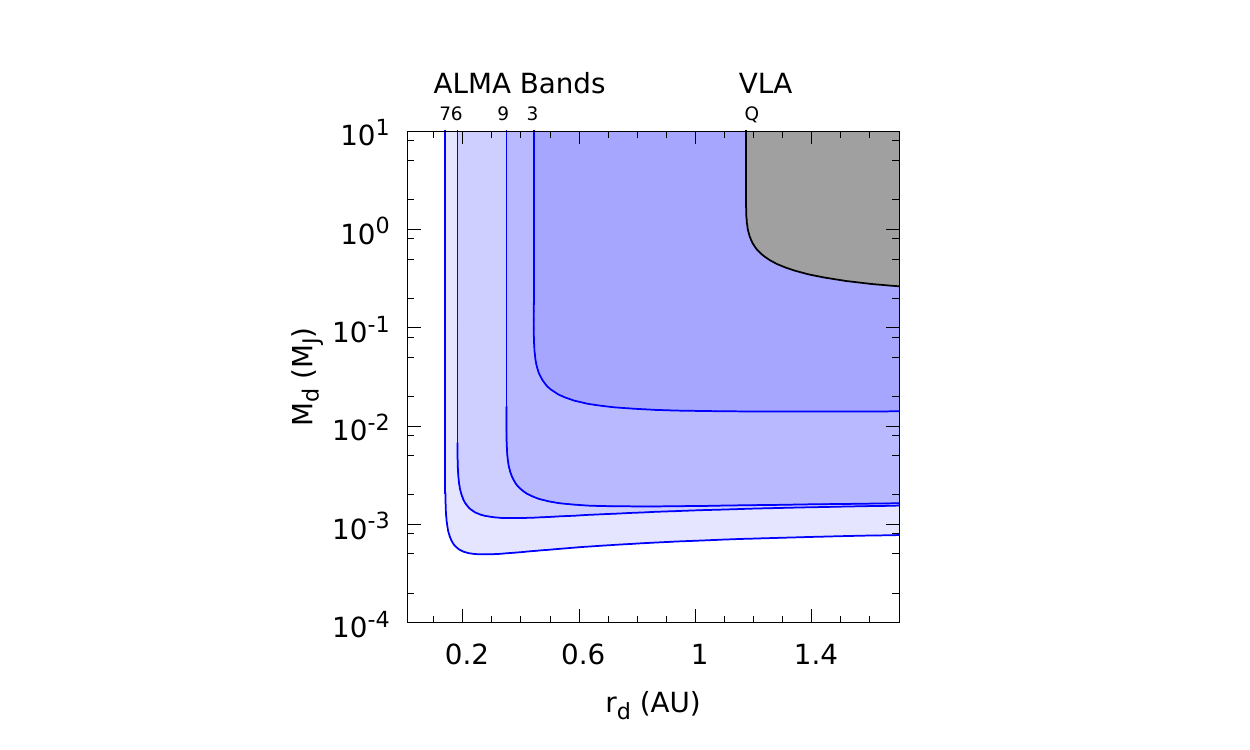}
\caption{\label{fig:alma_com} Parameter space for the mass and radius of LkCa~15~b's circumplanetary disk that can be probed
at more that  99.7\% confidence level in one hour of ALMA observations in the Band 3 (110 GHz), 6 (230 GHz), 7 (345 GHz), and 
9 (650 GHz).  The disk emission was calculated by assuming a mass accretion rate of $10^{-6} M_J$ yr$^{-1}$ 
and a dust opacity emissivity proportional to  $\nu^{1.0}$. The gray region corresponds to the region probed by our VLA observations 
and it is the same as in left panel of Figure 3. 
}
\end{figure}

Future ALMA observations will provide better constraints on the mass and radius 
of circumplanetary disks orbiting around LkCa~15, and, in general, within disks characterized 
by large dust depleted cavities.  ALMA full array will achieve in a few hours a sensitivity in disk mass 
100 times greater than that  obtained by VLA observations (Figure~\ref{fig:alma_com}) and 
an angular resolution sufficient to detect circumplanetary disks orbiting at a few AU from the 
central star at the distance of 140 pc.  Assuming a dust emissivity proportional to $\nu^{1.0}$,
ALMA band 7 observations ($\lambda=850$~$\mu$m) will achieve a mass sensitivity of 
$5 \times 10^{-4}$ $M_J$ in one hour of integration on source. Furthermore, ALMA observations in 
the other bands, although less sensitive, will allow the measurements of the spectral index 
of the continuum emission to study the evolution of solids, in a manner similar to that employed
 to study the grain size distribution in circumstellar disks  (see, e.g., the review by Testi et al. 2014).
 
 %\citep[see, e.g., the review by][]{Testi14}

%20~$\mu$Jy beam$^{-1}$ in 1 hr of integration on source at the wavelength 
%of 0.87 mm. Assuming a circumstellar disk spectral slope of 2.6, this corresponds to 
%a flux sensitivity of about 0.09 $\mu$Jy beam$^{-1}$ at 7~mm. For the same integration time, 
%ALMA observations will be therefore more than two order of magnitude more 
%sensitive than our VLA observations.

\acknowledgments
A.I., L.P., and J.M.C. acknowledge support from NSF award AST-1109334.
The National Radio Astronomy Observatory is a facility of the National Science 
Foundation operated under cooperative agreement by Associated Universities, Inc.

\bibliographystyle{apj}
%\bibliography{ref}
\bibliography{ref}

\end{document}